\documentclass[aps,prl,10pt,twocolumn,superscriptaddress]{revtex4-1}

\usepackage{times}
\usepackage{changes}
\usepackage{graphicx}
\usepackage{amssymb}
\usepackage{epstopdf}
\usepackage{amsmath}
\usepackage{color}
\usepackage{hyperref}
\usepackage{bbold}
\usepackage{float}
\usepackage{cancel}
\usepackage{url}
\usepackage{soul} 
\usepackage{dsfont} 

\usepackage[acronym]{glossaries}
\usepackage{physics}

\makeatletter

\@ifundefined{textcolor}{}
{
 \definecolor{BLACK}{gray}{0}
 \definecolor{WHITE}{gray}{1}
 \definecolor{RED}{rgb}{1,0,0}
 \definecolor{GREEN}{rgb}{0,1,0}
 \definecolor{BLUE}{rgb}{0,0,1}
 \definecolor{CYAN}{cmyk}{1,0,0,0}
 \definecolor{MAGENTA}{cmyk}{0,1,0,0}
 \definecolor{YELLOW}{cmyk}{0,0,1,0}
}

\graphicspath{{Images/}}

\hypersetup{
	pdfpagemode={UseOutlines},
	bookmarksopen,
	pdfstartview={FitH},
	colorlinks,
	linkcolor={black},
	citecolor={blue},
	urlcolor={blue}}
	
\usepackage{hyperref}
\hypersetup{
colorlinks=true,
linkcolor=black,           
citecolor=blue,           
filecolor=magenta,        
urlcolor=cyan
}

\makeatother

\begin{document}
\title{Simple Mitigation of Global Depolarizing Errors in Quantum Simulations}

\author{Joseph Vovrosh*}
\affiliation{Blackett Laboratory, Imperial College London, London SW7 2AZ, United Kingdom}
\author{Kiran E. Khosla}
\affiliation{Blackett Laboratory, Imperial College London, London SW7 2AZ, United Kingdom}
\author{Sean Greenaway}
\affiliation{Blackett Laboratory, Imperial College London, London SW7 2AZ, United Kingdom}
\author{Christopher Self}
\affiliation{Blackett Laboratory, Imperial College London, London SW7 2AZ, United Kingdom}
\author{M. S. Kim}
\affiliation{Blackett Laboratory, Imperial College London, London SW7 2AZ, United Kingdom}
\author{Johannes Knolle}
\affiliation{Department of Physics TQM, Technische Universit{\"a}t M{\"u}nchen, James-Franck-Stra{\ss}e 1, D-85748 Garching, Germany}
\affiliation{Munich Center for Quantum Science and Technology (MCQST), 80799 Munich, Germany}
\affiliation{Blackett Laboratory, Imperial College London, London SW7 2AZ, United Kingdom}

\begin{abstract}
 To get the best possible results from current quantum devices error mitigation is essential. In this work we present a simple but effective error mitigation technique based on the assumption that noise in a deep quantum circuit is well described by global depolarizing error channels. By measuring the errors directly on the device, we use an error model ansatz to infer error-free results from noisy data. We highlight the effectiveness of our mitigation via two examples of recent interest in quantum many-body physics: entanglement measurements and real time dynamics of confinement in quantum spin chains. Our technique enables us to get quantitative results from the IBM quantum computers showing signatures of confinement, i.e. we are able to extract the meson masses of the confined excitations which were previously out of reach. Additionally, we show the applicability of this mitigation protocol in a wider setting with numerical simulations of more general tasks using a realistic error model. Our protocol is device-independent, simply implementable and leads to large improvements in results if the global errors are well described by depolarization.
\end{abstract}

\maketitle
\section{Introduction} 
Quantum computers are becoming large enough ($\sim50-100$ qubits~\cite{preskill2018quantum}) to, in principle, allow demonstrations of their `Quantum Advantage'~\cite{boixo2018characterizing,arute2019quantum}. However, the actual amount of entanglement that can be generated in current devices is constrained by noise and errors, limiting their ability to solve complex problems such as quantum simulation. To address this, various \emph{error mitigation} strategies have recently been developed to counteract noise and boost the fidelity of experimental results.

Error mitigation differs from fault tolerance. Fault tolerant quantum computers will eventually be able to suppress errors by encoding quantum information in a redundantly large number of qubits as error correcting codes \cite{steane1996error,shor1995scheme}. In this way they will be able to execute arbitrarily deep circuits by the repeated application of active error corrections. 
Unfortunately, these encodings cannot be used in current devices as they require smaller hardware errors and larger numbers of qubits than are currently available. In contrast, error mitigation strategies are applied to unencoded physical qubits. Rather than actively correcting errors, they aim to estimate what the effect of the error was and infer the error-free result. 
This current stage in the development of quantum computers has been dubbed the noisy intermediate-scale quantum (NISQ) era \cite{preskill2018quantum} and is expected to last for the foreseeable future. 

\begin{figure}
    \centering
 \includegraphics[width=0.5\textwidth]{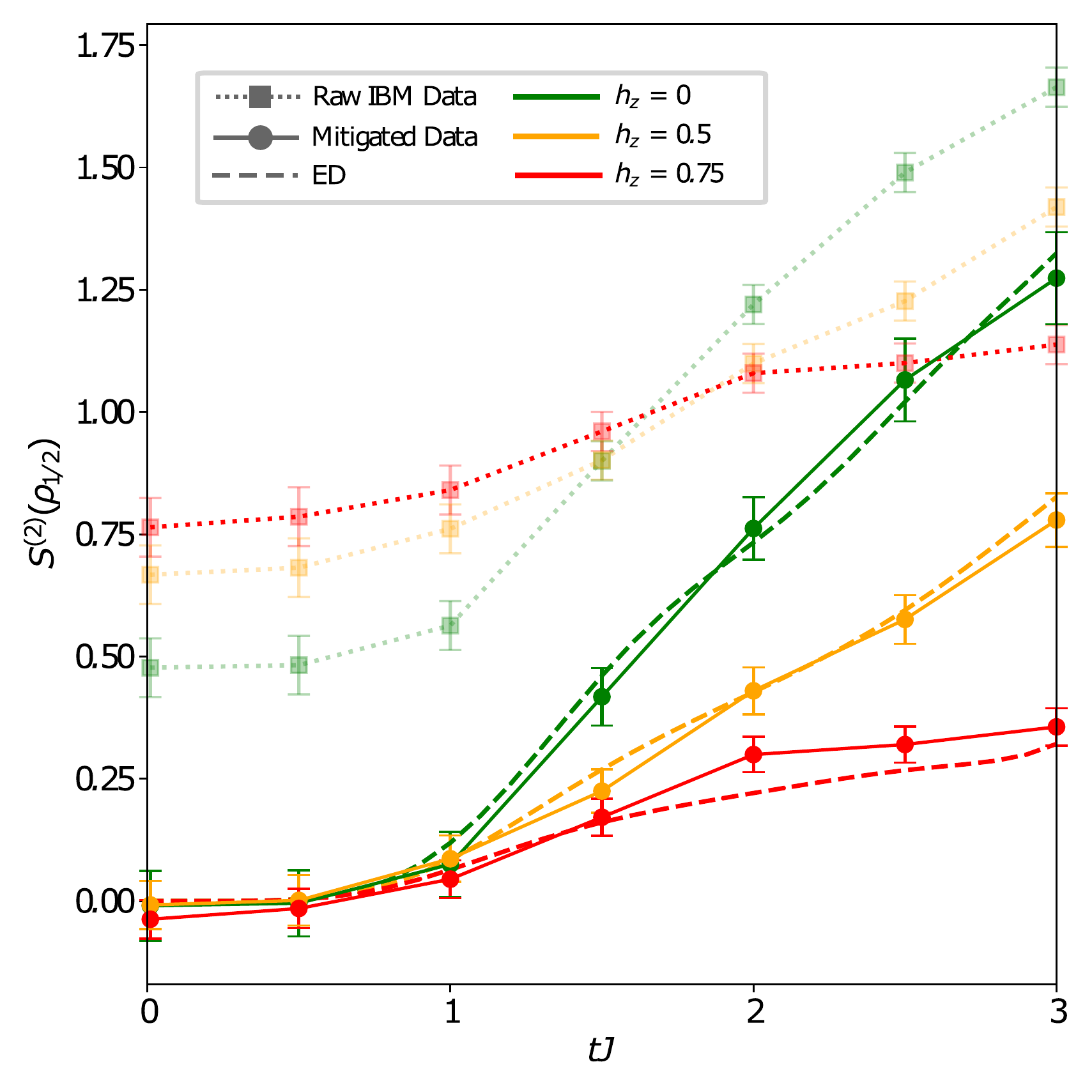}
    \caption{\textbf{Results from the IBM device \textit{Paris}}~\cite{paris} \textbf{of the second order R\'enyi entanglement entropy.} Quench dynamics before and after error mitigation of the second order R\'eyni entropy for the transverse field Ising model with varying longitudinal field strengths. Here, $J=1$, $h_x=0.5$ and $n=6$. Note, $800$ random unitaries were used to collect this data giving an uncertainty in the purity of $\delta \rho \sim O(10^{-2})$. Clearly the mitigation allows results to go from qualitative agreement to quantitative agreement with the results obtained numerically through exact diagonalization (ED).}
    \label{Fig1}
\end{figure}

In the past few years, error mitigation has been a thriving research direction as more and more quantum devices become available.  Error mitigation strategies typically address measurement errors~\cite{qiskit,jattana2020general, maciejewski2020mitigation,bravyi2020mitigating,chen2019detector,mcclean2020decoding,bonet2018low,sagastizabal2019experimental,tran2020faster} or the algorithms and gates employed for digital quantum simulation~\cite{huggins2019efficient,mcardle2019error,dumitrescu2018cloud,mcclean2017hybrid,colless2018computation,kandala2019error,otten2019accounting,otten2019recovering}. While some of these techniques have proven useful for certain quantum algorithms, their general success is hampered by the fact that they either rely on a high level of control of the quantum device itself, or are specifically designed for a given quantum simulation problem, i.e. exploiting specific symmetries~\cite{smith2019simulating}.

One promising direction is to employ machine learning algorithms~\cite{strikis2020learning,zlokapa2020deep,czarnik2020error} for error mitigation. Generally speaking, these methods train classical computers to predict the error seen in a quantum device and use the results to infer the error-free quantum simulations. While successful, these methods require a large increase in the classical computational overhead and are somewhat uncontrolled. The latter drawback is also true for popular protocols based on the idea of increasing errors in the device systematically and then extrapolating back to the zero error case~\cite{kandala2019error,li2017efficient, endo2018practical,temme2017error,gustafson2019quantum,giurgica2020digital,cai2020multi,he2020resource}. 
In general, how to tune the error rates varies from device to device and reliable fitting requires expert knowledge of the specific hardware.

Here, we propose a new protocol for gate error mitigation combining a raft of desirable features: it is easily implementable on any quantum device with little increase in workload, it is well rooted in a mathematical description of the errors, and it is suitable for any quantum algorithm of interest. Furthermore, as we show via specific examples, it can lead to to large improvements in the performance of quantum devices, for example see Fig. \ref{Fig1} for R\'enyi entropy results.  

The paper is organised as follows: First, we derive an ansatz for the density matrix resulting from the action of a noisy quantum simulator with global depolarizing errors. Following this, we explain how this ansatz can be used for a general error mitigation protocol. We then demonstrate its effectiveness by studying real time dynamics of the transverse field Ising model (TFIM) with a longitudinal field. In that context, some of the authors recently showed that signatures of confinement and entanglement spreading can be observed on the IBM quantum computer~\cite{vovrosh2020confinement}. With our new error mitigation protocol we are able to measure the meson masses of confinement induced bound states directly on the IBM device and can quantitatively measure entanglement spreading previously out of reach. Furthermore, we corroborate these results with classical simulations of error models taken from an IBMQ device in a more general setting to show the wide applicability of our mitigation scheme. Finally, we close with a discussion and outline future applications. 

\section{Error Mitigation Protocol} Error channels can be conveniently modelled through Kraus operators~\cite{nielsen2002quantum} which define a completely positive map on the density matrix
\begin{equation}
    \rho\rightarrow\sum_iK_i\rho K^{\dagger}_i \;\;\text{ such that }\;\;\sum_iK^{\dagger}_iK_i = 1.
\end{equation}
For a single qubit, choosing four Kraus operators to be proportional to the Pauli operators (identity included) defines a depolarizing channel. The proportionality constants are related to the individual error probabilities and must satisfy the identity constraint to preserve the trace of $\rho$.

Here, we concentrate on such depolarizing errors which are inevitably present in any digital quantum simulator platform and can be treated without much specific knowledge about the device performance (which also fluctuates over time). Moreover, our focus on depolarizing errors has the advantage of being treatable mathematically in a controlled way as detailed below. In fact, given a deep enough quantum circuit (with sufficient qubits) for self-averaging of incoherent errors to occur, this depolarizing error model is a good approximation to the physical errors in a device (more details are given in the supplementary material). Last but not least, the large improvement in the quality of results on the IBM device motivate our choice of depolarizing errors a posteriori.

An n-single qubit depolarizing error channel can be modelled via
\begin{equation}
    \mathcal{E}^{\otimes n}(\rho) = (1-p)^n\rho + \sum_{\alpha\in[x,y,z]}\sum_{j=1}^n(1-p)^{n-1}\frac{p}{3}\sigma_\alpha^j\rho\sigma_\alpha^j + ...
    \label{Eq:2}
\end{equation}
where $\mathcal{E}$ is the error channel, $p$ is the probability of an error occurring for each qubit (assumed to be equal for each qubit, and for each Pauli error) and `...' indicates higher-order terms corresponding to errors on multiple qubits \cite{nielsen2002quantum}. One important feature of this mathematical formulation is that the second term, which describes the depolarizing error, commutes with any unitary operator. Consequently, the error on the $i^{th}$ qubit in a quantum circuit with purely depolarizing errors is
\begin{equation}
    \mathcal{E}^{i}(\rho) = (1-p_i)\rho + p\Tr_i[\rho]\otimes\frac{\mathcal{I}_i}{2}
    \label{Eq:3}
\end{equation}
in which $\rho$ is the density matrix, $\mathcal{I}_i/2$ is the maximally mixed (i.e. completely depolarized) state for the $i^{th}$ qubit, $\Tr_i$ is the partial trace over the $i^{th}$ qubit and $p_{i}$ is the error on the $i^{th}$ qubit. 

Instead of dealing with all combinations of single qubit errors, we approximate the total error channel of Eq.~\eqref{Eq:2}, under the assumption of symmetric depolarization Eq.~\eqref{Eq:3}, as an effective depolarizing channel on the \emph{entire} quantum state 
\begin{equation}
    \rho = (1-p_{\mathrm{tot}})\rho_{\mathrm{exact}} + p_{\mathrm{tot}}\frac{\mathcal{I}^{\otimes n}}{2^n}
    \label{Eq:4}
\end{equation}
where the effective \textit{total} error probability is $p_\mathrm{tot}$. In principle $p_{tot}$ is well approximated by $\prod_i(1-p_i)$, however, we do not make that identification here. We will soon show that $p_{\mathrm{tot}}$ can be measured directly on the device. The simplicity of this ansatz allows it to be easily calculated. It has already been shown to be useful when mitigating measurement error \cite{vermersch2018unitary}, however we later demonstrate that it is powerful tool for mitigating global depolarizing errors which do not themselves originate from local depolarizing errors. Furthermore, we later numerically demonstrate that this ansatz does not rely on the assumption of single- and two-qubit depolarizing errors.

The many partial traces over single qubits, which would have conserved some coherence in the remaining qubits, have been replaced by the maximally mixed state $\mathcal{I}^{\otimes n}/2^n$ over the global quantum state, destroying all coherence. We stress that even though this may not be a good approximation for a single layer of qubit errors, it becomes a reasonable approximation for the error channel of a many layer, many qubit circuit.
Eq.~\eqref{Eq:4} is our basic ansatz for an effective error model after a many-layered unitary circuit and $\rho_{\mathrm{exact}}$ is the exact density operator without noise (see Supplementary Material for more details). With this ansatz one can analytically calculate the effect of errors on a measured observable, $\hat{O}$, via 
\begin{equation}
\langle \hat{O}\rangle = (1-p_{\mathrm{tot}})\Tr[\hat{O}\rho_{\mathrm{exact}}] + \frac{p_{\mathrm{tot}}}{2^n}\Tr\big[\hat{O}\big].
\label{Eq:5}
\end{equation}

Our approach estimates $p_{tot}$ in order to apply error mitigation. We propose and test two approaches for finding $p_{tot}$, the first based on estimating the purity of the final state and the second based on studying specific observables. To estimate the purity, we employ the recent protocol for obtaining the trace of the reduced density matrix squared $\Tr[\rho_A^2]$ via randomised measurements~\cite{van2012measuring,elben2018renyi}, where $A$ is a subspace of the full density matrix. This randomised measurement scheme has been successfully implemented in trapped ion quantum simulators~\cite{brydges2019probing} and recently by some of us on the IBM quantum computer~\cite{vovrosh2020confinement}. We stress the present mitigation is a far simpler task compared to inverting the quantum error channel to tomographically reconstruct the error free quantum state $\rho$.

As current quantum devices initialise systems in pure states that are then manipulated with unitary transformations, $\Tr[\rho^2]$ over the full Hilbert space should lead to a result that is identically one. However, since the noisy implementation of quantum circuits will in general deviate from unitarity, after a given quantum circuit is run on a quantum processor this will generally not be the case. Instead, with Eq.\eqref{Eq:4} we expect that
\begin{equation}
    \Tr[\rho^2] = (1-p_{\mathrm{tot}})^2 + \frac{p_{\mathrm{tot}}(1-p_{\mathrm{tot}})}{2^{n-1}} + \frac{p^2_{tot}}{2^{n}}
    \label{Eq:6}
\end{equation}
Using the fact that $\rho_{\mathrm{exact}}$ is pure.
Now, given that the left hand side, $\Tr[\rho^2]$, can be measured directly on the device~\cite{vovrosh2020confinement,brydges2019probing}, this quadratic equation can be solved to obtain the total error $p_{\mathrm{tot}}$ \cite{vermersch2018unitary,elben2020cross}. We stress that correlated errors in quantum circuits do not increase entropy and thus $p_{\mathrm{tot}}$ obtained via this method should really be understood as the \textit{global depolarizing} error probability. While this mitigation approach does not address coherent errors, it could additionally be combined with other techniques such as twirling \cite{bennett1996mixed,bennett1996purification,knill2008randomized,emerson2007symmetrized}.

Therefore, with $p_{\mathrm{tot}}$ extracted and $\langle \hat{O}\rangle$ measured the only unknown quantity in Eq. (\ref{Eq:5}) is the desired error-free observable $\Tr[\hat{O}\rho_{\mathrm{exact}}]$. Note, we assume that $\Tr[\hat{O}]$ can be calculated which for most practical cases should be the case, e.g. see our application examples below. 

Putting all steps together, we finally obtain our general error mitigation protocol:
\begin{enumerate}
    \item Prepare the quantum state of interest by running a quantum circuit and measure $\Tr(\rho^2)$, e.g. via randomized measurements~\cite{elben2018renyi,brydges2019probing,vovrosh2020confinement}.
    \item Use the results to obtain values for $p_{\mathrm{tot}}$ via Eq. (\ref{Eq:6}).
    \item Prepare the quantum state again and measure the desired observable $\langle \hat{O}\rangle$. \footnote{As errors on the quantum devices can vary over time (see Ref. \cite{smith2019simulating}), the determination of $p_{\mathrm{tot}}$ has to be recalibrated accordingly.}
    \item Use Eq. (\ref{Eq:5}) with the measured value of $p_{\mathrm{tot}}$ to obtain the desired $\Tr[\hat{O}\rho_{\mathrm{exact}}]$.
\end{enumerate}

An alternative approach to estimating $p_{tot}$ is to consider specific observables whose expectation values are known. For example, consider the time dynamics of a system in which our quantum circuit approximates the time evolution operator $U(t)=\exp{-iHt}$ where we wish to measure $\langle O(t) \rangle$ for a range of times. We can tune the circuit such that ${t\ast(E_{max})=\epsilon<<1}$ where $E_{max}$ is the largest eigenvalue of the Hamiltonian (shifted so $E_{min} = 0$), so that our quantum circuit now approximates the identity operation. Assuming that $\langle O(t=0) \rangle$ is known, we can use the measurements from the quantum device and Eq.(\ref{Eq:5}) to obtain $p_{tot}$. Note, for infinite dimensional Hamiltonian with unbounded eigenvalues, the approximation is slightly more subtle, but does not apply to qubits. We show that, while this method is more efficient, it does not discriminate between coherent and incoherent errors in the purity.

Our protocol can be applied to essentially any quantum circuit and quantum simulation device. In the following, we choose a representative example from condensed matter physics as a first application. We showcase the effectiveness of our technique by presenting previously unobtainable results for confinement and entanglement dynamics in spin chains. 

\section{Application to spin chain confinement} An ideal testing ground for NISQ devices is that of quench dynamics in spin-$\frac{1}{2}$ systems. A global quantum quench is a sudden change to the systems Hamiltonian, which induces non-equilibrium dynamics. Already one dimensional spin chains can show a wide variety of physical phenomena of interest, for example confinement of domain wall excitations~\cite{kormos2017real,liu2018confined,vovrosh2020confinement}, quantum many-body scars~\cite{turner2018weak,moudgalya2018exact,ho2019periodic,mark2020unified,shibata2020onsager}, or novel fracton excitations~\cite{nandkishore2019fractons,pai2020fractons,sala2020ergodicity}. All of these show up in the time evolution which is challenging to simulate on classical computers as the Hilbert space grows exponentially $2^n$ with the number of spins $n$. As spin-$\frac{1}{2}$ systems directly map onto physical qubits, quantum computers are ideally suited for studying the rich physics of spin chains. Recently, first digital quantum simulation results have been reported~\cite{cervera2018exact, zhukov2018algorithmic,francis2019quantum, smith2019simulating, smith2019crossing, vovrosh2020confinement} but in order to obtain results which are out of reach by classical simulations and to probe non-trivial quantum many body physics better error mitigation techniques are needed.

We concentrate on the one-dimensional TFIM with an additional longitudinal field given by the following Hamiltonian
\begin{equation}
    H = -J\bigg[\sum_i\sigma_i^{z}\sigma_{i+1}^{z} + h_x\sum_i \sigma_i^{x}+ h_z\sum_{i} \sigma^{z}\bigg],
    \label{Eq:7}
\end{equation}
where $J$ is the Ising exchange of nearest neighbour spins $\sigma_i$ and $h_{x/z}$ are the relative strengths of the transverse and longitudinal fields respectively. For $h_z = 0$ the TFIM can be solved exactly via Jordan-Wigner transformation and its fermionic excitations are related to free domain wall motion. When turning on the longitudinal field, $h_z\neq0$, a confining potential between these fermions is introduced. The attraction between fermions grows linearly with their separation, reminiscent of quark confinement in QCD. The result is the formation of `mesonic' bound states of domain wall excitations.

\begin{figure}
    \centering
    \includegraphics[width=0.45\textwidth]{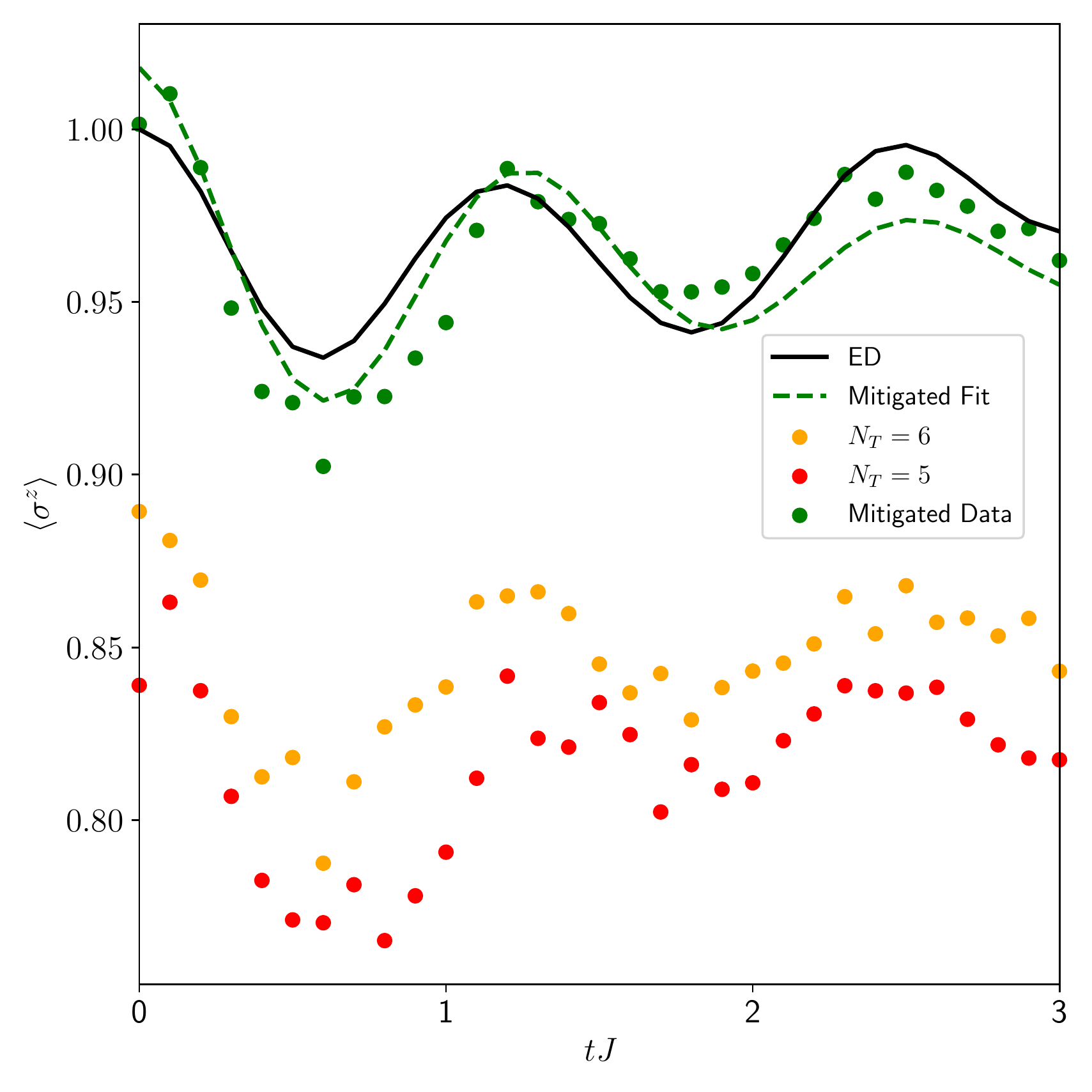}
    \caption{\textbf{Results from the IBM machine before and after mitigation.} Quench dynamics of local magnetization from Trotterized time evolution of the TFIM with longitudinal field before and after error mitigation. Here, $J=1$, $h_x=0.5$, $h_x=0.75$ and $n = 7$. Data is shown for, $N_t$, the number of trotter steps $N_t = 5,6$. Results go from qualitative agreement to quantitative agreement. After fitting a cosine function (dashed green) to the mitigated data (green dots) the dominant frequency is clearly captured by the IBM device, for details see Supplementary Material. Note, more details of the circuit composition of the evolution operator can be found in \cite{vovrosh2020confinement}}
    \label{Fig2}
\end{figure}

The confining potential between fermions induces non-ergodic behaviour~\cite{james2019nonthermal}, which manifests itself through persistent oscillations in the local magnetization, $\langle\sigma_i^\alpha\rangle$ ($\alpha \in \{x,y,z\}$) and a slowing down of the entanglement spreading. 

\subsection{Example 1: Measuring Meson Masses}
The frequencies of the oscillating magnetization can be mapped directly to the energy of the domain wall bound states~\cite{liu2018confined}, which have a large overlap with chosen initial states. These so called \textit{meson masses} are defined as the energy difference between the lowest excited states and the ground state. 

Recent work using a trapped ion quantum simulator to simulate the long-range TFIM model showed how, by choosing a variety of initial states, the meson masses can be measured through the persistent oscillations of local magnetization~\cite{tan2021domain}. Previous attempts to perform a similar measurement on a digital quantum computer have failed for the short ranged TFIM, Eq. (\ref{Eq:7}), because the results are too noisy to resolve the smaller amplitude of oscillations~\cite{vovrosh2020confinement}. As our main result, we show that our new error mitigation enables us to obtain the meson masses from the IBM device. 

We are mainly interested in the time dependence of the local magnetization, which further simplifies with $\Tr [\sigma^\alpha ]=0$ in Eq. (\ref{Eq:5}) to
\begin{equation}
    \langle\sigma_i^\alpha\rangle =  (1-p_{\mathrm{tot}})\langle\sigma_i^\alpha\rangle_{\mathrm{exact}}.
    \label{Eq:8}
\end{equation}
The time dependence can be calculated by applying a quantum circuit from a Trotterisation of the time evolution operator (for details of the implementation and quantum circuits see Refs.~\cite{smith2019simulating,vovrosh2020confinement}). Increasing the number of trotter steps, $N_t$, leads to a deeper circuit and the ensuing increase in errors results in a peculiar dampening of the magnetization dynamics which can be removed via our error mitigation.
In Fig. \ref{Fig2} we display the results of the local magnetization dynamics. Our mitigation technique not only allows us to measure the main oscillation frequency on the IBM device but in fact to do so with quantitative agreement with the analytical results (the latter is explained in Ref. \cite{vovrosh2020confinement}). 

\begin{figure*}
    \centering
    \includegraphics[width=1.\textwidth]{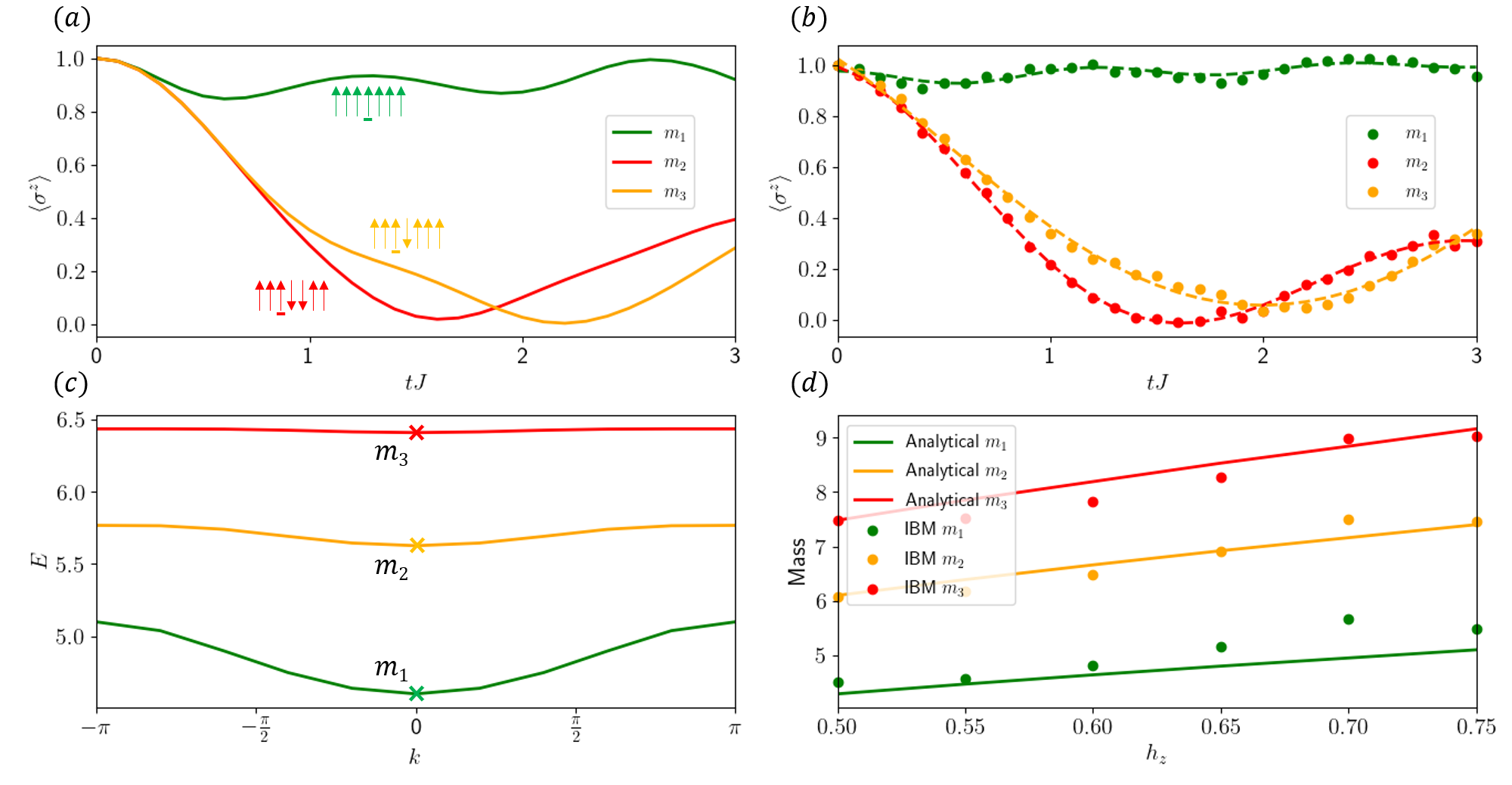}
    \caption{\textbf{Results from the IBM device \textit{Toronto}}~\cite{toronto} \textbf{of meson masses.} (a,b) The quench dynamics of the $z$-axis local magnetization is shown for different initial states. Here, $J=1$, $h_x=0.75$, $h_x=0.75$ and $n=7$. (a) Results calculated via exact diagonalizion. (b) Mitigated results from the IBM device. Here, clear dominant oscillations are extracted that quantitatively agree with the analytically derived values. (c) An illustration of how the masses are extracted from the analytically derived energy levels~\cite{vovrosh2020confinement}. (d) A comparison of the masses obtained from the IBM device and the analytically derived values for varying $h_z$ showing this quantitative agreement. Note, more details of the circuit composition of the evolution operator can be found in \cite{vovrosh2020confinement}}
    \label{Fig3}
\end{figure*}

We note that an additional simplification can be used for the local magnetization for added efficiency --- no additional measurement of $\Tr[\rho^2]$ was required because the result of $\langle\sigma^z\rangle$ is known at $tJ = 0$ and thus $p_{\mathrm{tot}}$ can be inferred from the corresponding measurement on the IBM device after running the time evolution circuit for $tJ \approx 0$. This big simplification avoids the costly randomized measurement scheme and should be generally applicable for observables whose value is known at $tJ = 0$. However, as this does not distinguish correlated noise from the uncorrelated noise assumed in our ansatz, it is possible for over-estimations in the mitigation to occur, resulting in non-physical results. An example of this is seen in Fig. \ref{Fig2} at time $tJ = 0.1$ in which the local magnetization is measured to be greater than unity.

We have obtained a range of results with different number of trotter steps. Under the assumption that $(1-p_{\mathrm{tot}})$ scales as $(1-p_T)^{N_T}$, where $p_T$ is the error in one trotter step, we can extrapolate the results back to the error-free case, see green data points in Fig. \ref{Fig2}. 
We are then in a position to suppress the noise to a level which enables us to extract different meson masses on the IBM device by a basic fit of the main oscillation frequency. In Fig. \ref{Fig3} we show the data for the first three masses obtainable by starting from different initial states, see insets. Here, 7 spins are mapped onto 5 qubits resulting in a circuit with $5N_T+5$ single qubit gates and $8N_T$ CNOT gates with $N_T =5,6$. Remarkably, we find quantitative agreement between the mitigated results and the theoretical predictions for the scaling of the meson masses with the transverse and longitudinal fields~\cite{vovrosh2020confinement}. 
    
\subsection{Example 2: Entanglement Spreading}
As a second example of we study the suppression of half chain entanglement entropy spreading after the spin chain quench. For $h_z=0$ the entanglement entropy is expected to increase linearly from the ballistic spreading of free fermionic excitations~\cite{fagotti2008evolution}. However, with a non-zero longitudinal field this growth is suppressed in a characteristic fashion due to confinement~\cite{kormos2017real}. To observe this we have implemented the randomized measurement protocol of Ref. \cite{brydges2019probing} on the IBM device for measuring the second order R\'eyni entanglement entropy. In Ref. \cite{vovrosh2020confinement} two of us obtained qualitative agreement for entanglement dynamics of six spins compared to an exact diagonalization (ED) calculation, but for a quantitative agreement a large shift of the results was needed. 

By using the ansatz in Eq. (\ref{Eq:4}) we can see that the effect of global depolarizing errors on measurements of $\Tr[\rho_A^2]$, where $A$ is a subsystem in consideration, is
\begin{equation}
\begin{aligned}
    \Tr[\rho_A^2] =& (1-p_{\mathrm{tot}})^2\Tr[\rho_{A,exact}^2] \\ &+ \frac{p_{\mathrm{tot}}(1-p_{\mathrm{tot}})}{2^{n_A-1}} + \frac{p_{tot^2}}{2^{n_A}}.
    \label{Eq:9}
    \end{aligned}
\end{equation}
If $p_{\mathrm{tot}}$ is known, $\Tr[\rho_{A,exact}^2]$ can be extracted and the second order R\'enyi entropy measurement is calculated via
\begin{equation}
    S^{(2)}(\rho_{\mathrm{exact}}) = -\log_2(\Tr[\rho_{A,exact}^2]).
\end{equation}
Fig. \ref{Fig1} shows how this mitigation protocol eliminates the error in the second order R\'enyi entropy results. With our error mitigation protocol and Eq. (\ref{Eq:9}) we now obtain quantitative agreement with ED results for six spins and can account for the large shift of the results.

\section{Applications to more general examples} 
	
To demonstrate that our method works more generally, we simulate example tasks of measuring the expectation value of different operators. We choose a layered, brickwork circuit consisting of a layer of single qubit rotations followed by a layer of entangling CNOTs --- the same circuit structure as for Trotterized evolution (see Supplementary material). However, instead of time parameterized Trotter evolution, we simply choose random single qubit angles, corresponding to different states, and therefore different expectation values.

In order to obtain a larger amount of data to support our mitigation technique, we numerically simulate the layered circuits with noisy gates. Qiskit's circuit simulator allows one to simulate arbitrary gate-based noise models by specifying independent Kraus operators for each single- and two-qubit gates. Using this noisy simulator, we can directly test how gate-level non-depolarizing errors can result in an effective global circuit-level depolarizing error, and how well our mitigation technique works in the presence of local non-depolarizing errors. To simulate a realistic noise model, we take the Kraus operators directly from the \textit{ibmq\_santiago} backend \cite{santiago}. This error model goes beyond the single gate depolarizing assumption, by including (asymmetric) thermal relaxation.

In order to test local, non-local, single, and many Pauli-string operators, we have chosen the following operators: single $Z$ operator (local, single term), TFIM Hamiltonian (local, many terms), Random Pauli string (non-local, single term), and the molecular Hamiltonian of a H$_4$ (non-local, many terms). Here many 'terms refers' to the number of non-commuting Pauli strings that must be measured to construct the expectation value, and locality refers to only containing single and two-qubit Pauli strings. Note the four hydrogen Hamiltonian (see Supplementary for details) is chosen as a simple molecular test case with a large Hilbert space, and where we can ignore complications from freezing out orbitals. 

\begin{figure}
    \centering
    \includegraphics[width=.5\textwidth]{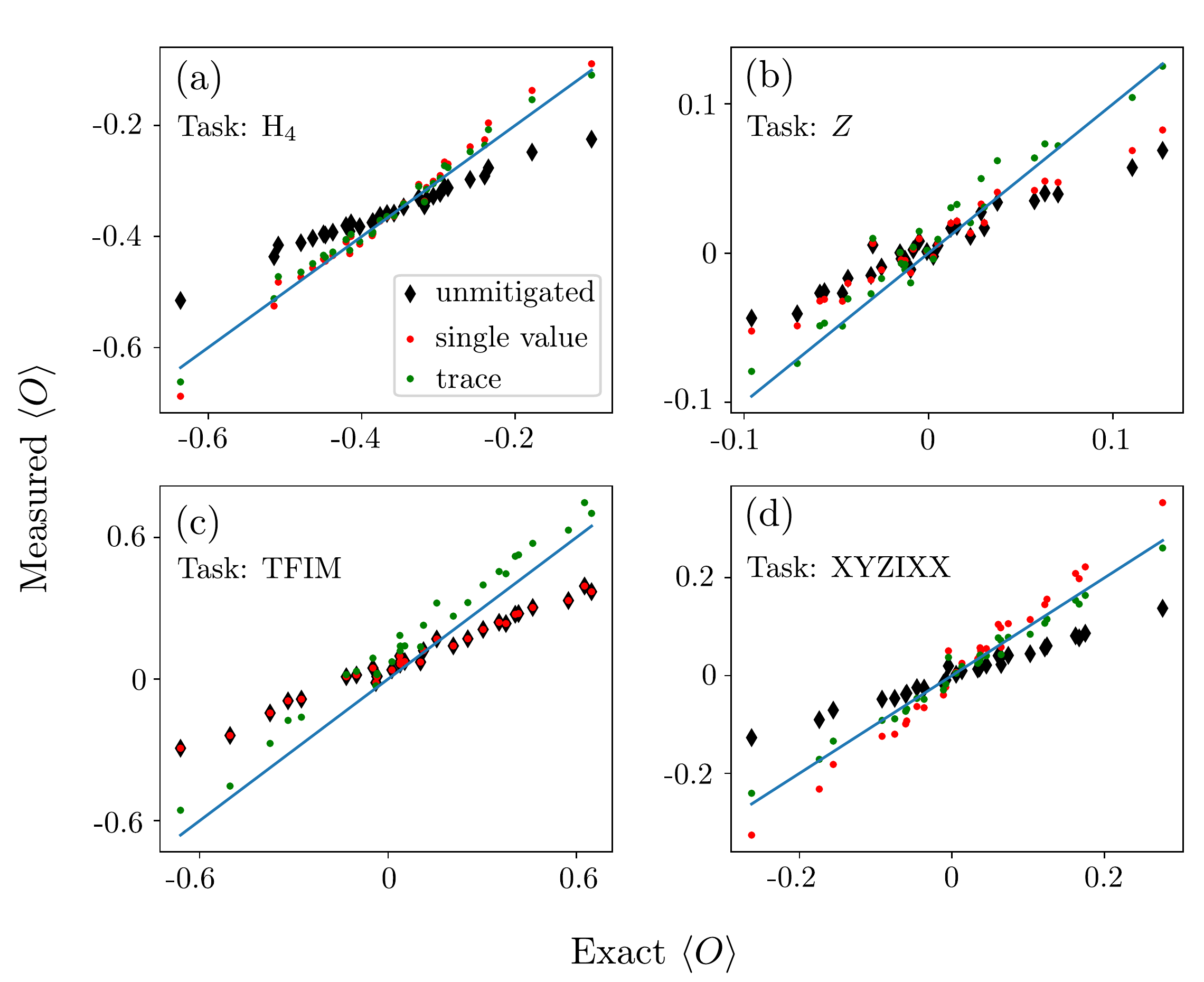}
    \caption{\textbf{Error mitigation simulations with non-depolarizing errors.} Unmitigated (black diamonds), trace mitigated (green circles) and single operator mitigated (red circles) expectation values for different operators, with the solid blue line showing perfect mitigation. The numerical simulation uses the general error model taken from \textit{ibmq\_santiago}. Expectation values are evaluated with respect to parameterized brick-like circuits with (CNOT) depth 18 (see supp for details), and thirty random parameter points per plot. Operators are (a) Hamiltonian of 4 linear-equidistant hydrogen atoms (6 qubits), (b) Single $Z$ Pauli term (8 qubits), (c) TFIM Hamiltonian (6 qubits) and (d) Single $XYZIXX$ string (6 qubits). These operators are chosen to span local/global Pauli strings, and contain single/many non-commuting decompositions.}
\label{fig:sims}
\end{figure}

Figure~\ref{fig:sims} shows unmitigated and mitigated expectation values for each operator, and clearly demonstrates the added value of purity measurements (see Supplementary for simulation details). For expectation value mitigation (i.e. using a single expectation value to calibrate $p_{tot}$), a single extra circuit is needed. For trace mitigation (using Eq.~\eqref{Eq:9} to calibrate $p_{tot}$) uses 500 extra circuits to find $p_{tot}$; these are split into five groups to reduce the bias. Once $p_{tot}$ is calibrated for a given circuit depth, this single value is used for error mitigation of the same circuit structure, but for different single qubit parameter angles. 

As the main results we find that our {\it global} depolarising assumption is an effective error channel for the whole circuit, and, crucially, it does not require each single- and two-qubit errors to be themselves purely depolarising.

\section{Discussion} In this work we have proposed an error mitigation technique which is simple to implement but which retains the mathematical rigour of more complicated techniques. Our protocol is directly applicable to any quantum simulation whose measurements are basic expectation values, as exemplified by our results for the time evolution of the local magnetization in spin chain dynamics. Though our error mitigation method may not replace full state tomography, we expect it to be useful in measuring more complicated quantities beyond simple observables, as corroborated by our results for the entanglement entropy. An interesting avenue of future research will be an application to variational quantum eigensolver (VQE) problems~\cite{kandala2017hardware} and other quantum circuits.

The computational overhead of implementing our protocol is dominated by the cost of evaluating $\Tr(\rho^2)$ on the quantum device. The number of corresponding measurements is $N_uN_m$ where $N_u$ is the number of randomised unitaries and $N_m$ is the number of random measurements. Within the randomised scheme of Ref.~\cite{elben2018renyi} it grows exponentially with the system size~\cite{brydges2019probing}, which poses a potential problem for large quantum computers but is easily feasible for currently available NISQ devices. 
Alas, in some cases the costly randomized measurement scheme can be avoided entirely, i.e. in our benchmark example of spin chain dynamics a single measurement of a local observable whose exact value is known was sufficient. This should be true generally for quantum circuits which can be tuned to be close to the identity. However, this scalable simplification for obtaining $p_{\mathrm{tot}}$ potentially faces the problem that it includes correlated errors which makes the randomised measurements preferable as long as it is feasible.

The fact that our basic assumption of global depolarizing errors leads to such large improvements in results is in itself remarkable. The basic conclusion is that the total error on the IBM device is close to a global depolarizing error at least for our choice of problems. We note our error ansatz, Eq.~\eqref{Eq:4} is an approximation even for single qubit depolarizing errors, let alone the more complex channels that are no-doubt present in physical devices. Nevertheless the ansatz works remarkably well for correcting physical errors of large/deep circuits and is a no-lose addition to quantum simulation protocols. We suggest this is because the depolarizing channel is a good approximation \emph{per gate} in a many gate quantum circuit, even if this approximation breaks down for single layers of gates. Of course, our mitigation falls short of accounting for large coherent or correlated errors and it will be a worthwhile endeavour to think about a controlled extension of our basic ansatz Eq. (\ref{Eq:4}) for the density matrix of a NISQ device. In particular, whether we can extend it to incorporate other aspects of error channels~\cite{cai2020multi}.

As an application of our error mitigation we have presented previously unobtainable quantitative results for confinement and entanglement dynamics of a quantum spin chain. We have been able to extract the first meson masses of confinement induced bound states and observed the corresponding halting of entanglement spreading. 
In that context, an ambitious next step would be to use the error mitigation to extend times that can be simulated on a quantum computer to probe the confinement induced slow-thermalisation~\cite{liu2018confined} or to see meson scattering events that have recently been predicted~\cite{Surace2020scattering,Karpov2020Spatiotemporal,milsted2020collisions}. 

In general, we expect that our simple error mitigation brings us closer to exploiting the quantum advantage of available NISQ devices for real world practical applications. 

\begin{acknowledgments}
{\it Acknowledgements.---} 
We are grateful for discussions with Hongzheng Zhao, Adam Smith and Peter Haynes. We acknowledge the Samsung Advanced Institute of Technology Global Research Partnership, travel support via the Imperial-TUM flagship partnership and the use of IBM Quantum services for this work. The views expressed are those of the authors, and do not reflect the official policy or position of IBM or the IBM Quantum team. This work is also supported by the UK Hub in Quantum Computing and Simulation, part of the UK National Quantum Technologies Programme with funding from UKRI EPSRC grant EP/T001062/1.
\end{acknowledgments}

\bibliographystyle{apsrev4-1}
\bibliography{Mitigation}
\section{Supplementary Information}
\subsection{Justification of a global depolarizing Error Model}

Here we justify why such a simple global depolarizing error model is applicable. For simplicity we consider single qubit gate errors, modelled as depolarizing errors, before moving onto multiqubit gates. The error channel for an operator acting only on qubit $i$ is, $\mathcal{E}_i[\rho]= (1-p) U_i\rho U_i^\dagger + p \mbox{Tr}_i[\rho]\otimes \mathcal{I}_i/2$ where $U_i$ acts locally on qubit $i$. That is with probability $1-p$ the gate was successfully applied, and with probability $p$ the qubit suffered a completely depolarizing error, and ends up in the maximally mixed state $\mathcal{I}/2 = \mbox{diag}[1/2, 1/2]$, thereby destroying quantum correlations between qubit $i$ and the reset of the state. This error model does not commute with entangling gates (e.g. consider the reduced state $\mbox{Tr}_i[\rho]$ depends on the entanglement between qubit $i$ and the rest of the state). While applying single gates to multiple different qubits commute, the resulting expression quickly becomes difficult to track, $\mathcal{E}_i\circ\mathcal{E}_j[\rho] = (1-p_i)(1-p_j)U_iU_j\rho U_j^\dagger U_i^\dagger + p_i(1-p_j)\mbox{Tr}_i[U_j\rho U_j^\dagger]\otimes \mathcal{I}_i/2 + p_j(1-p_i)\mbox{Tr}_j[U_i\rho U_i^\dagger]\otimes \mathcal{I}_j/2 + p_ip_j\mbox{Tr}_{ij}[\rho]\otimes \mathcal{I}_i/2 \otimes \mathcal{I}_j/2$. 

Our first step is to approximate a layer of single qubit gates \footnote{Assuming many qubits, but not necessarily all qubits have gates applied} as 
\begin{eqnarray}
\mathcal{E}_{\mathrm{layer}} = (1-p')U\rho U^\dagger + p'\mathcal{I}/2^n,
\label{eq:map_layer}
\end{eqnarray}
for $n$ qubits. We don't set $(1-p')\neq \prod_i (1-p_i)$, but rather introduce an effective $p'$ to account for components in the total sum that have some reasonable overlap with $U\rho U^\dagger$. E.g. for each $j$, the state $\mbox{Tr}_j[U\rho U^\dagger]\otimes \mathcal{I}_j/2$ may have a large overlap with $U\rho U^\dagger$, (and to a lesser extent $\mbox{Tr}_{ij}$ for each $i,j$), and we'd like to include such contributions without worrying about combinatorial multiplicity, or the state-dependent details of all the partial traces. We can follow the same steps to arrive at Eq.~\eqref{eq:map_layer} for entangling gates $U_{ij}$ once we note that the 2-qubit depolarizing error channel be written as $\mathcal{E}_{ij}[\rho]= (1-p) U_{ij}\rho U_{ij}^\dagger + p \mbox{Tr}_{ij}[\rho]\otimes \mathcal{I}_{ij}/4$, where $\mathcal{I}_{ij}/4 = \mbox{diag}[1,1,1,1]/4$ is the maximally mixed two qubit state. 

We can construct the quantum channel for the full circuit. The non-unitary part of Eq.~\eqref{eq:map_layer} commutes with itself over different layers meaning the a full circuit error channel is approximated as 
\begin{eqnarray}
\mathcal{E}_{\mathrm{circ}}[\rho] = (1-p_{\mathrm{tot}})U_{\mathrm{circ}}\rho U_{\mathrm{circ}}^\dagger + p_{\mathrm{tot}}\mathcal{I}_2^{\otimes n}/2^n
\label{eq:map_circ}
\end{eqnarray}
where $(1-p_{\mathrm{tot}}) = \prod_l (1-p_l')$ (product over effective layer $p'$'s) holds due to commuting non-unitary terms. While this is not equal to $\prod_g (1-p_g)$ (product over gate error probabilities), the latter is a good zeroth-order estimate of this term. 

Finally we note the total map Eq.~\eqref{eq:map_circ} unravels to an effective error model for single (i.e. one or two qubit) gates, given by: $\mathcal{E}_i[\rho] = (1-p_i'')U_i\rho U_i^\dagger + p_i'' \mathcal{I}_2^{\otimes n}/2^n$. Note the all-or-nothing description of the map: either the gate is implemented perfectly, or the \emph{entire} quantum state is destroyed, not just part of the state as is the case for $\mbox{Tr}_i[\rho]\otimes \mathcal{I}_i/2$ etc. While this seems quite an unphysical model for a single gate, it is a good model of the effective gate error per gate in a multi-gate circuit.

\subsection{Cosine Fitting}

In order to extract the frequencies from the local magnetization results presented in Fig. \ref{Fig2} and Fig. \ref{Fig3}, a function that contained a singular cosine was used, namely,
\begin{equation}
    \langle\sigma^z(t)\rangle \sim A e^{-dt} \cos(\omega t) + c_1t + c_2.
\end{equation}
Here, $A$ is the amplitude of the oscillations, $D$ allows for decay of the amplitude, $\omega$ is the oscillation frequency and $c_1$ and $c_2$ give a time dependent shift. While this function can not capture long term behaviour of $\langle\sigma^z\rangle$, it is able to extract out the dominant oscillations for short time dynamics.

\subsection{Statistical Errors Introduced by the Error Mitigation}
The error when calculating $p_{tot}$ from $\Tr(\rho^2)$ is given by
\begin{equation}
    \delta p_{tot} = \pm(1-p_{tot})\frac{\delta\Tr(\rho^2)}{2(2^{n}-\Tr(\rho^2))},
\end{equation}
where $\delta p_{tot}$ is the error in $p_{tot}$ and $\delta\Tr(\rho^2)$ is the error in the measurement of $\Tr(\rho^2)$. Note that $\delta\Tr(\rho^2)$ can be calculated by resampling e.g. Jackknife or Bootstrap \cite{brydges2019probing,vovrosh2020confinement}. Thus, when using the proposed error mitigation on an observable $\hat{O}$ this results additional uncertainty on top of the measurement errors. More precisely,
\begin{equation}
    \delta\langle\hat{O}\rangle = \pm\Bigg( \frac{\delta_{meas}\langle\hat{O}\rangle}{1-p_{tot}} + \frac{\delta p_{tot}\big(\langle\hat{O}\rangle - 2^{-n}\Tr(\hat{O})\big)}{(1-p_{tot})^2}\Bigg),
\end{equation}
where $\delta\langle\hat{O}\rangle$ is the error in the mitigated value of $\langle\hat{O}\rangle$ and $\delta_{meas}\langle\hat{O}\rangle$ is the error in the measurement of $\langle\hat{O}\rangle$. Note, as the local magnetisation results presented in this paper bypassed the measurement of $Tr(\rho^2)$, the error contributing to $\delta p_{tot}$ is just shot noise. Thus, $\delta \langle\sigma_i^z\rangle$ is of the on the order $\sim O(10^{-3})$ and is not displayed in Fig. \ref{Fig2} or Fig. \ref{Fig3}.

\subsection{Numerical simulations}
\label{sec:app_numerical_sims}
All numerical simulations are performed using qiskit \cite{qiskit} with the santaiago error model (as of 12$^th$ March 2021). For the hydrogen Hamiltonian, openfermion \cite{McClean2020Jun} was used to compute the orbital integrals (for a row of four H-atoms with 0.1~nm inter-atomic distance), and then map the problem to qubits using the symmetry conserving Bravyi-Kitaev  transformation~\cite{Bravyi2017Jan}, finally the $\mathds{Z}_2$ symmetries (corresponding to particle spin) are removed, reducing the number of qubits by two. For the TFIM, the Hamiltonian was reduced to two non-commuting Pauli strings from which the total Hamiltonian can be evaluated. Finally the single $Z$, and $XYZIXX$ (chosen randomly) operators were measured trivially. We do not directly account for State-Preparation and Measurement (SPAM) errors, but we note that we can (and do) include SPAM errors as contributing to $p_{tot}$ in our channel ansatz. 

We numerically simulate circuits consisting of layers of parameterized \emph{u3} gates, generic single-qubit rotation gates with 3 Euler angles, followed by an entangling layer. The circuit depth is defined by the number of entangling (CNOT) layers. There is a final \emph{u3} layer before measurement. The parameterized circuit now creates a state $\rho(\theta)$, and varying $\theta$ changes the state and corresponding operator expectation values. For the numerical simulations we consider how our mitigation performs for different depth circuits, and for the four tasks shown in the main text. For each depth the calibration $p_{tot}$ must be done independently, as $p_{tot}$ is an \textit{effective} parameter for the whole quantum channel and cannot simple be calculated \textit{a priori} for different depths. 

\textit{Calibration.---} Two independent methods of calibrating $p_{tot}$ are considered. Firstly, via single expectation value: direct simulation of a circuit with fixed parameters, which is then compared to the error free expectation value to find $p_{tot}$. Secondly, via trace: from directly measuring $\mathrm{Tr}[\rho^2]$ and estimating $p_{tot}$ via Eq.~\eqref{Eq:9}. 

For single expectation calibration, a single fixed parameter circuit was used to find $\langle O \rangle$ (for each operator $O$ considered). This result was compared to the error free simulation of the exact same circuit to estimate $p_{tot}$. To evaluate these expectation values 40960 shots were used for each non-commuting term in the decomposition of the given operator into Pauli strings. For trace mitigation, an extra 500 circuits are used to estimate $p_{tot}$ using 8196 shots per circuit. This was broken up into five lots of 100 circuits to estimate $p_{tot}$ at five different (random) circuit parameterizations. For each parameterization, the method of Ref.~\cite{elben2018renyi} is used to find five independent estimates of $p_{tot}$, which are then averaged to reduce the bias in the estimate of $p_{tot}$ for the circuit. 

\begin{figure*}
	\centering
	\includegraphics[width=\textwidth]{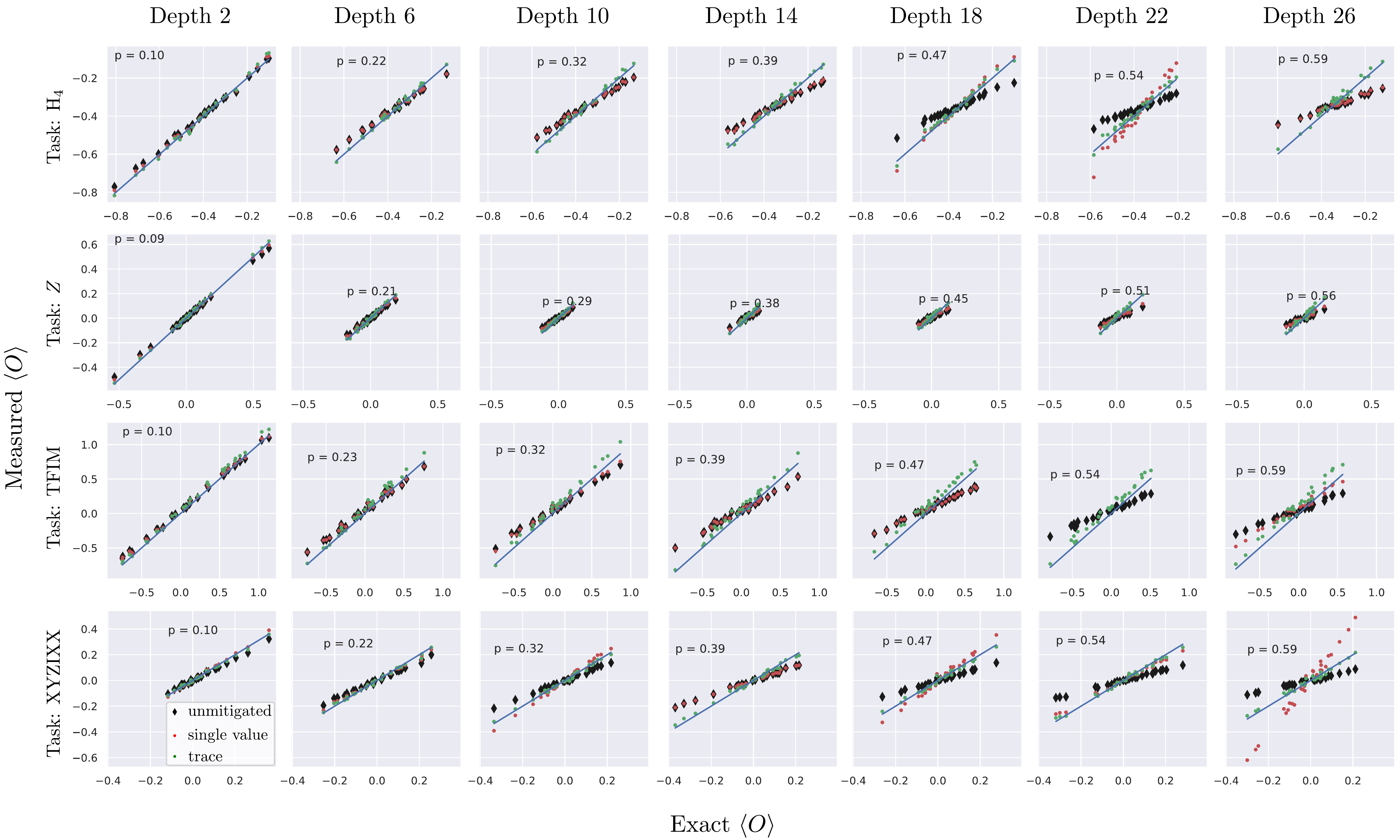}
	\caption{\textbf{Complete mitigation simulations with non-depolarizing errors.} Scaling of our mitigation schemes (circles) with circuit depth for four different operator expectation values. Single operator (red circles) show some improvement, while trace mitigated (green circles) show a robust improvement over the unmitigated results. As in the main text, numerical simulation uses the \textit{ibmq\_santiago} error model, and circuits are layers \emph{u3} and CNOT, with the depth equal to the number of entangling layers. Note there are some combinations (e.g. TFIM at depth 14 etc) where the single expectation value mitigation found $p_{tot} = 1$, and therefore did not improve results. Likewise there are other combinations (e.g. TFIM depth 22) where the single expectation value mitigation is omitted because it gave an unphysical estimate of $p_{tot} <0$. Inset numbers show $p_{tot}$ as estimated for the trace estimation.}
	\label{fig:all_sims}
\end{figure*}

\textit{Mitigation.---} Once $p_{tot}$ is estimated either from the single value or trace measurement, it is used to mitigate further simulated expectation values. New expectation values are obtained by varying the parameters in the circuit, while keeping the circuit structure fixed. For each depth and each task, thirty different (parameterized) circuits are simulated, and mitigation applied. Figure \ref{fig:all_sims} shows the mitigation results for different circuit depths. While the full trace-estimation is robust showing a significant improvement in every case, the single expectation value mitigation is more variable. This is due to the vanishing expectation value of Pauli strings for random circuits, which in turn increases the effect of shot noise when inverting Eq.~\ref{Eq:8}. We note this shot noise problem is absent when calibrating a Pauli string (or single operator) with a (noise free) unit expectation value, as in the main text. 

\end{document}